\documentclass[12pt]{article}
\usepackage[margin=1in]{geometry}
\usepackage{bm}
\usepackage{setspace}
\usepackage{graphicx}
\usepackage{amssymb}
\usepackage{amsmath}
\usepackage{epstopdf}
\usepackage{algorithm}
\usepackage{amsfonts}
\usepackage{rotating,color}
\usepackage{subfigure}
\usepackage{natbib}
\usepackage{multirow}
\usepackage{titlesec, textcase}
\usepackage{longtable}
\usepackage{bbm}
\usepackage{rotating}

\allowdisplaybreaks[3]

\titleformat{\section}{\bf\large\center\uppercase}{\thesection}{1em}{}

\def\bI{{\boldsymbol I}}

\def\cO{{\cal O}}

\def\bx{{\boldsymbol{x}}}

\def\bX{{\boldsymbol X}}

\def\by{{\boldsymbol y}}

\def\bzero{{\boldsymbol 0}}

\def\veps{\varepsilon}

\def\argmin{\mathop{\rm argmin}}
\def\real{\mathop{{\rm I}\kern-.2em\hbox{\rm R}}\nolimits}

\newcommand{\pkg}[1]{$\mathsf{ #1}$}

\def\sgn{\hbox{sgn}}

\def\hbeta{\hat{\beta}}

\def\bH{{\boldsymbol H}}

\def\bzero{{\boldsymbol 0}}

\def\bx{{\boldsymbol x}}
\def\bX{{\boldsymbol X}}
\def\bM{{\boldsymbol M}}

\def\bbeta{\boldsymbol \beta}
\def\hbbeta{\hat{\boldsymbol \beta}}

\def\lam{\lambda}

\def\bzero{{\bf 0}}
\def\bSigma{\boldsymbol\Sigma}

\def\bhbeta{\hat{\boldsymbol \beta}}
\def\cM{\mathcal{M}}

\definecolor{grey1}{gray}{0.5}
\definecolor{grey2}{gray}{0.3}

\def\boxit#1{\vbox{\hrule\hbox{\vrule\kern6pt\vbox{\kern6pt#1\kern6pt}\kern6pt\vrule}\hrule}}

\newtheorem{example}{Example}
\begin{document}

\title{Modified Cross-Validation for Penalized High-Dimensional Linear Regression Models
\thanks{Yi Yu is Research Associate, Statistical Laboratory, Cambridge University, U.K. CB30WB (Email: y.yu@statslab.cam.ac.uk). Yang Feng is Assistant Professor, Department of Statistics, Columbia University, New York, NY 10027 (Email: yangfeng@stat.columbia.edu).  }
}
\author{Yi Yu and Yang Feng}
\date{}
\maketitle

\begin{abstract}
In this paper, for Lasso penalized linear regression models in high-dimensional settings, we propose a modified cross-validation method
for selecting the penalty parameter. The methodology is extended to other penalties, such as Elastic Net. We conduct extensive simulation studies and real data analysis to compare the performance of the modified cross-validation method with other methods. It is shown that the popular $K$-fold cross-validation method includes many noise variables in the selected model, while the modified cross-validation works well in a wide range of coefficient and correlation settings. Supplemental materials containing the computer code are available online.
\end{abstract}


\section{Introduction}

To analyze high-dimensional data, variable selection is a popular tool. \cite{Tibshirani1996} proposed Lasso, which is the $\ell_1$ penalty, or equivalently \cite{ChenDonoho1994} proposed Basis Pursuit. Later, Elastic Net variants \citep{ZouHastie2005} and non-convex penalties such as SCAD \citep{FanLi2001} and MCP \citep{Zhang2010} were proposed and widely used over the years. All of these variable selection procedures proved to have good theoretical properties.

Besides, developing efficient algorithms for calculating the solution path of the coefficient vector as tuning parameter varies is of great importance. A vast literature on calculating the path for penalized linear regression is available. Among these, least angle regression (LARS) \citep{EfronHJT2007}, or homotopy \citep{Osborne2000}, Local Quadratic Approximation (LQA) \citep{FanLi2001}, Local Linear Approximation (LLA) \citep{ZouLi2008}, Penalized Linear Unbiased Selection (PLUS) \citep{Zhang2010}, and coordinate descent methods \citep{Fu1998, FreidmanHHHT2007}  gained popularity these days.

After getting a path of solutions from the foregoing mentioned methods, users still need to pick one estimator from the path with different penalty levels controlled by the tuning parameter. As it turns out, selecting the optimal tuning parameter is both important and difficult. There has been a line of research on using information-type criteria to select the tuning parameter. \cite{Tibshirani1996} used generalized cross-validation (GCV) style statistics, and \cite{EfronHJT2007} used $C_p$ style statistics. \cite{ZouHT2007} derived a consistent estimator for the degree of freedom of Lasso, and plugged it into the $C_p$, AIC, and BIC criteria. But for Lasso estimators in high-dimensional setting,  from simulation experience, all these traditional methods tend to over-select, due to the bias introduced by shrinkage. \cite{ChenChen2008} proposed extended-BIC, by adding an extra term with respect to $p$ to the information criterion. Motivated by generalized information criterion (GIC) proposed by \cite{Nishii1984}, \cite{ZhangEtal2010} extended GIC to a more general scenario, which can handle non-convex penalized likelihood estimation. In \cite{WangEtal2009}, they conjectured that the traditional BIC-type criterion tends to over-select in high-dimensional scenarios and proposed a modified BIC-type criterion.

Another popular family of methods for selecting the tuning parameter is cross-validation, which is a data-driven approach. A majority of theoretical work has been done for cross-validation in the classical linear regression models. For example, leave-one-out cross-validation (CV(1)) is shown to be asymptotically equivalent to AIC, the $C_p$, the jackknife, and the bootstrap \citep{Stone1977, Efron1983, Efron1986}. \cite{Shao1993} gave rigorous proof of the inconsistency of CV(1) for linear regression model, meanwhile he provided the proper sizes of construction and validation set in leave-$n_v$-out cross-validation (CV($n_v$)), under which cross-validation achieves the model selection consistency. \cite{ZhangP1993} studied multifold cross-validation and $r$-fold cross-validation in linear regression models. It turned out both methods tend to select more variables than the truth under certain technical conditions. For several popular packages in \pkg{R} for Lasso, e.g. \pkg{lars} \citep{EfronHJT2007}, \pkg{glmnet} \citep{FriedmanHT2010}, \pkg{glmpath} \citep{ParkHastie2007}, $K$-fold cross-validation is still the default option. Researchers have realized that the regular cross-validation in high-dimensional settings tends to be too conservative in the sense that it selects a majority of false positives. As mentioned in \cite{ZhangHuang2008}, the theoretical justification of cross-validation based penalty parameter choice is unclear for model selection purposes. Cross-validation is also mentioned in \cite{Meinshausen}, where relaxed Lasso is proposed, which includes LARS-OLS \citep{EfronHJT2007} as a special case. In that paper, the author conjectured by using $K$-fold cross-validation, relaxed Lasso estimator is model selection consistent. The tuning parameter selection problem also exists for other type of variable selection methods, e.g.,  the adding-noise approach in \cite{LuoEtal2006} and \cite{WuEtal2007}. 

In this paper, we aim to develop a new cross-validation approach for selecting tuning parameter for high-dimensional penalized linear regression problems. It is noteworthy that we are not proposing a new variable selection technique, rather, the goal is to study and improve the variable selection performance for the existing tuning parameter selection methods. The contribution of the paper is two-fold. (1) A thorough investigation on several popular cross-validation methods is conducted, and they are shown to be inconsistent via simulation. (2) A modified cross-validation criterion is provided, which is shown to have better performance in terms of model selection and prediction under high-dimensional settings. 

The rest of the paper is organized as follows. We introduce the model setup and fix notations in Section \ref{sec-model}, and propose the modified cross-validation criterion in Section \ref{sec-mcc}. Extensive simulation studies, including various simulation settings and comparisons to the existing methods, and real data analysis are conducted in Sections \ref{sim} and \ref{sec-data}, respectively. A short discussion is presented in Section \ref{sec-discussion}.

\section{Model Setup}\label{sec-model}

Given $n$ observation pairs $( \bx_i,y_i)$, $i = 1, \cdots, n$, we consider the linear regression model
\begin{align*}
y_i = \bx_i'\bbeta + \veps_i,
\end{align*}
where $\bx_i$'s and $\bbeta$ are $p$-dimensional vectors, with $p \gg n $. $\veps_i$'s are iid random variables with mean 0 and variance $\sigma^2$.

Denote $\bX=(\bx_1,\cdots,\bx_n)'$ as the $n\times p$-dimensional design matrix and $\by=(y_1,\cdots,y_n)'$ as the response vector. We employ notations $\|\cdot\|$ and $\|\cdot\|_1$ as the $\ell_2$ and $\ell_1$ norms of a vector, respectively; $\|\cdot\|_0$ as the number of non-zero entries of a vector; and $\bbeta$ as the true $\bbeta$, satisfying $\|\bbeta\|_0 = d_0 < n$. The oracle set $\{j:~\beta_j\neq 0\}$ is denoted as $\cO$.
To analyze this high-dimensional problem, we adopt the popular Lasso estimator \citep{Tibshirani1996}, i.e.,
\begin{align*}
\bhbeta({\lam}) \equiv \arg\min_{\bbeta} \frac{1}{2n} \|\by-\bX\bbeta\|^2 + \lambda \|\bbeta\|_1.
\end{align*}
Using any path algorithm, we can get a solution path $\bhbeta(\lam)$ when $\lam$ changes. Notice, the \emph{starting point} of this paper is that we have a collection of estimators on hand with the goal of choosing the optimal one among them. For the Lasso estimators, it is equivalent to study how to choose the optimal tuning parameter $\lam$. Both $\lam$ and $\hat{\bbeta}$ are functions of $n$, but to keep the notations simple, we omit the subscript $n$ throughout the paper.

Following the notation system developed in \cite{Shao1993}, we denote the model corresponding to tuning parameter $\lam$ as $\cM_{\lam}$ and divide the collection of models $\{\cM_{\lam}, \lam>0\}$ into two disjoint categories. Category I includes the models which miss at least  one important variable, i.e., false negative number is greater than 0 (FN $>0$). Category II includes the models with all the important variables, i.e.,  FN = 0. The optimal model $\cM_{\lam_*}$ is defined to be the model of the smallest size among Category II models, i.e., the most parsimonious candidate model that contains all the important variables. If there are more than one models satisfying these two properties, we define the one with the smallest $\lam$ as the optimal one, considering the shrinkage brought in by $\lam$. Here, $\lam_*$ is the tuning parameter of the optimal model. Now the goal is to find $\cM_{\lam_*}$ and the corresponding $\lam_*$.

By exploiting the cross-validation approach, the main idea is to repeatedly split the data into construction and validation sets, fit the model for the construction data, and evaluate the performance of the fitted model on the validation set. The prediction performance over different splits are averaged and the average can represent the predictability of the model. One can then choose the best model according to the predictability measure of different models. In \cite{Shao1993}, the average is taken over the same models estimated from different splits. However, for the case of cross-validation in Lasso estimators, it is not always possible to perform the same averaging because one will generally get different models from different splits even if the same $\lam$ values are used. Instead of averaging the performance for the same model, we measure the performance of the models corresponding to the specific $\lam$ value.

For cross-validation, denote $s$ as a subset of $\{1, \cdots, n\}$ containing $n_v$ integers for validation, and $(-s)$ as  its complement containing $n_c$ integers, where $n_v+n_c=n$. Considering the sub-samples and  their associated sub-models will appear later, we denote $\cM_{(-s), \lam}$ as the model constructed from sample $(-s)$ given $\lambda$. The following notations are used throughout the paper.  The subscripts are based on the submatrices for the corresponding construction sets and the tuning parameter $\lambda$.
\begin{align*}
&\bX_{s, \lam} = (x_{ij}),~ i\in s, ~ j\in \cM_{(-s), \lam}; \quad  \bX_{(-s), \lam} = (x_{ij}),~ i\in (-s), ~ j\in \cM_{(-s), \lam}; \\
& \bX_{\cdot,\lam} = (x_{ij}),~ i = 1, \cdots, n,~ j\in \cM_{\lam}; \quad \bX_{\cdot, \lam(s)} = (x_{ij}),~ i = 1, \cdots, n,~ j\in \cM_{(-s), \lam};\\
& \bH_{s,\lam} = \bX_{s, \lam}\left(\bX_{\cdot, \lam(s)}'\bX_{\cdot, \lam(s)}\right)^{-1}\bX_{s, \lam}';
 \end{align*} $\by_s = (y_i,~ i\in s)'$; denote $\bhbeta_{(-s), \lam}$ as the Lasso estimator of $\bbeta$ under $\cM_{(-s), \lam}$.

\section{Modified Cross-Validation}\label{sec-mcc}

To deal with the over-selection issue of the traditional cross-validation in Lasso penalized high-dimensional variable selection, a new cross-validation method is proposed. Instead of developing a new variable selection technique, we would rather say the goal here is to investigate and improve the existing cross-validation methods. 

\subsection{Algorithm}
First, we describe a generic cross-validation algorithm for the Lasso estimators in the linear regression.

\begin{enumerate}
\item[S1.] Compute the Lasso solution path with the whole dataset. A sequence of solutions $\hbbeta(\lam)$ are generated with corresponding penalty level $\lam$'s.
\item[S2.] Randomly split the whole dataset into construction dataset (size $n_c$) and validation dataset (size $n_v$) $b$ times, compute the Lasso solution path for each construction dataset with the $\lam$ sequence in S1.
\item[S3.] For each split, use the corresponding validation dataset to calculate the values of the criterion function (to be specified)  for each path, and average over the paths with the same $\lambda$.
\item[S4.] Find the $\hat{\lam}$ with the smallest average criterion value, then fit a linear regression for the model $\cM_{\hat{\lam}}$. This linear regression estimator is the final estimator.
\end{enumerate}
The sequence of solutions $\hbbeta(\lam)$ mentioned in S1 is generated from certain Lasso path algorithm, such as \pkg{glmnet} used in subsequent simulations. Since the goal is to choose the optimal model from a collection of candidate models, the path generation method is not specified in the algorithm description. In S4, considering the bias caused by Lasso procedure, a further linear regression on the selected variable set is conducted after variable selection. The algorithm involves several parameters $n_c$, $n_v$, $b$, and the criterion function used in S3. We are interested in how these parameters and the criterion function affect the final estimator and which are the best ones. 

\subsection{Criterion Function}
In this subsection, we study the choice of criterion function used in S3. In the traditional cross-validation, the criterion function to be minimized is as follows.
\begin{align*}
  \Gamma_0(\lam)=\frac{1}{n_v}\|\by_s - {\hat \by}_{(-s), \lam}\|^2,
\end{align*}
where ${\hat \by}_{(-s), \lam} = \bX_{s,\lam}{\hat\bbeta}_{(-s),\lam}$ represents the predicted value on the subset $s$ using the Lasso estimate based on the data $(-s)$ when the penalty level is $\lam$.

Via numerical experience, researchers realized traditional cross-validation based on $\Gamma_0$ tends to select many false positives. This is mainly caused by the bias issue of the Lasso penalty. For convenience, we assume that all the matrix inversions appearing in the paper are well defined. This is to say, that for any subset $A \subset \{1,\cdots,p\}$ with small enough size appearing in this paper, $\bX_{A,A}'\bX_{A,A}$ is of full rank \citep{Zhang2010}. Now, we introduce the new cross-validation criterion \emph{Exactly Modified Cross-validation Criterion (EMCC)}, which is defined as
\begin{align}\label{newCri}
\Gamma_1(\lam) = \frac{1}{n_v}\|\by_s - {\hat \by}_{(-s), \lam}\|^2
 - \frac{\lam^2 n_c^2}{n_v}\bM_{s,\lam}'\bM_{s,\lam},
\end{align}
where $\bM_{s,\lam}=\bX_{s, \lam}\left(\bX'_{(-s), \lam} \bX_{(-s), \lam}\right)^{-1}\left(\sgn(\bhbeta_{(-s), \lam})\right)$, and
%
%
$\sgn(\cdot)$ represents the sign function.

Let $d_{(-s), \lam}$ be the model size of the current Lasso estimator ${\hat\bbeta}_{(-s),\lam}$. If the covariates are standardized and independent, we have 
$E(\bX'_{(-s), \lam} \bX_{(-s), \lam})=n_c\bI_{d_{(-s),\lam}}$ and $E(\bX_{s, \lam}'\bX_{s, \lam})=n_v\bI_{d_{(-s),\lam}}$. 
Now, if we replace the sample covariance matrices by their population versions, EMCC can be approximately reduced to the following simple form
\begin{align}\label{newCriOr}
\Gamma_2(\lam)=\frac{1}{n_v}\|\by_s - {\hat \by}_{(-s), \lam}\|^2 - \lam^{2}d_{(-s), \lam},
\end{align}
which is called \emph{Modified Cross-validation Criterion (MCC)}.  It is clear that MCC can be easily calculated by the knowledge of the current penalty level $\lam$ and the current model size $d_{(-s),\lam}$.

Briefly speaking, the EMCC criterion is designed to remove the systematic bias introduced by the shrinkage. Now, we give the detailed rationale behind it. Define $\hat\bbeta_{\cdot, \lam}$ as the subvector of $\hat\bbeta(\lam)$ restricted on $\cM_{\lam}$; to make notationally consistent, we put $\cdot$ in the subscript to indicate the estimator is derived using the whole dataset. Define $\tilde{\bbeta}_{\cdot, \lam}$ as the least squares estimator (LSE) for the model $\cM_{\lam}$, i.e.,
\begin{align*}
\tilde{\bbeta}_{\cdot,\lam} = \argmin_{\bbeta \in \mathbb{R}^{d_{\cdot, \lam}}} \frac{1}{2n}\|\by - \bX_{\cdot, \lam}\bbeta\|^2,
\end{align*}
where $d_{\cdot, \lam}$ represents the model size for penalty level $\lam$ using all the sample.
Correspondingly, ${\tilde \by}_{(-s), \lam} = \bX_{s,\lam}{\tilde\bbeta}_{(-s),\lam}$.
Recall the solution to the KKT conditions is the unique minimizer of the original optimization problem,
\begin{align}\label{eq::KKT}
\begin{cases}
\frac{1}{n}\bx_{j}'(\by - \bX\bhbeta) = \lam \sgn(\hat\beta_j), \quad \hbeta_j\neq 0; \\
 \frac{1}{n}\left|\bx_{j}'(\by - \bX\bhbeta)\right| \le \lam, \quad \hbeta_j = 0.
\end{cases}
\end{align}
Rearranging the terms in \eqref{eq::KKT}, we have the implicit expression, $\bhbeta_{\cdot, \lambda} = \left(\bX'_{\cdot, \lambda}\bX_{\cdot, \lambda}\right)^{-1}(\bX_{\cdot, \lambda}'\by - n\lambda \sgn(\bhbeta_{\cdot, \lambda}))$.

The Lasso prediction error based on the construction dataset $(-s)$ and validation dataset $s$ can be analyzed via inserting the prediction based on corresponding LSE, i.e. $\tilde{\by}_{(-s),\lam}$. We have,
\begin{align*}
 \|\by_s - \hat{\by}_{(-s),\lam}\|^2  &= \|\by_s - \tilde{\by}_{(-s),\lam}\|^2 
 +  \|\hat{\by}_{(-s),\lam} - \tilde{\by}_{(-s),\lam}\|^2 +(\by_s - \tilde{\by}_{(-s),\lam})'(\hat{\by}_{(-s),\lam} - \tilde{\by}_{(-s),\lam})\\
 &\equiv\|\by_s - \tilde{\by}_{(-s),\lam}\|^2 
 + (I) +(II). 
\end{align*}
The expectation of $(II)$ equals 0, 
and $(I)$ can be simplified as $\lam^2 n_c^2/n_v\bM_{s, \lam}'\bM_{s, \lam}$ via straightforward matrix operations. So in order to get rid of the systematic bias, EMCC is derived from subtracting  $(I)$ on both sides.


We call the cross-validation methods using \eqref{newCri} and \eqref{newCriOr} \emph{Exactly Modified Cross-Validation} and \emph{Modified Cross-Validation}, respectively. It is expected that there is a tradeoff between the accuracy and computational efficiency. These two criteria will be compared with the traditional cross-validation and other methods  in Section \ref{sim}.

\subsection{Data Splitting Strategy}

In this subsection, we study the choices of $n_c$, $n_v$ and $b$. CV($n_v$) with $n_v/n \to 1$, and $n_c \to \infty$ as $n\to \infty$
works well in model selection for fixed dimensional linear regression model \citep{Shao1993}. For notational simplicity, if without extra explanation,  by CV($n_v$) we mean CV($n_v$) with $n_v/n \to 1$, and $n_c \to \infty$, as $n\to \infty$. The simplification is also applied to other cross-validation methods to be introduced later.

 Here, to improve computational efficiency in high-dimensional settings,  instead of carrying out the calculation for all different splits when $n_v>1$ (which is of the order $p\choose n_v$),  we apply Monte Carlo method to split the dataset, by randomly drawing (with or without replacement) a collection of $b$ subsets of $\{1, \cdots, n\}$ with size $n_v$ and selecting the model with minimum average criterion function value over all splits. The Monte Carlo cross-validation was also considered in \cite{PicardCook1984} and \cite{Shao1993}.

We would like to point out that the EMCC calculation is equivalent to the LSE of the model sequence on the solution path, which is closely related to the LARS-OLS \citep{EfronHJT2007}. As suggested by an anonymous referee, a brief theoretical comparison is conducted below, with more emphasis on the computational issues in sequel. For Lasso penalized estimation problem, the important goal of model selection  is usually substituted by tuning parameter selection. However, in the cross-validation algorithm, different splits could lead to completely different model sequences even under the same tuning parameter sequence (see \cite{FengYu2013} for a detailed discussion). As a result, the theoretical analysis will involve the investigation of the average performance for different models, where the rigorous argument for either the proposed methods or LARS-OLS  are very involved. 

Considering this, we compare the proposal methods with LARS-OLS regarding data splitting strategy under an ideal scenario -- the induced model sequence under the same tuning parameter sequence are the same for different splits as that for the whole dataset.  For this ideal setting, \cite{Shao1993} proved that under mild conditions, cross-validation based on least squares estimators are consistent when $n_c/n\to 0$; otherwise, the cross-validation procedure fails to be consistent. Noteworthy, this holds for our proposed algorithms, while $n_c/n = (K -1)/K > 0$ in the LARS-OLS algorithm.

\subsection{Extensions}

Before concluding this section, it is worthwhile to highlight the main idea of  (E)MCC for Lasso penalized linear regression. Motivated by  the over-selection phenomenon in cross-validation procedure caused by shrinkage, (E)MCC is developed via removing the shrinkage. Based on the linear regression on the subset of covariates, leave-$n_v$-out data splitting strategy is used. Back to the comparison to LARS-OLS, due to the simplicity of the Lasso penalty, we have the approximated version MCC, which does not require any matrix operations when the solution path is available.  By calculating the LSE for the model sequence of a given solution path, the EMCC idea can be easily extended to other popular penalties, such as SCAD \citep{FanLi2001}, Elastic Net \citep{ZouHastie2005}, MCP \citep{Zhang2010}, among others.  The algorithm can be extended to  a general penalty by replacing all the solution path calculations in S1 and S2 by the ones using a general penalty, and use the following new criteria function in S3
\begin{align}\label{CriExt}
\Gamma_3(\lam)=\frac{1}{n_v}\|\by_s - {\tilde \by}_{(-s), \lam}\|^2. 
\end{align}
A simulation example for Elastic Net is available in Section \ref{sim}.

\section{Simulation}\label{sim}

In this section, we study the performance of EMCC/MCC based CV($n_v$). In Example \ref{ex1}, we introduce the basic setup of the simulation with various correlation settings, including an extension to Elastic Net. In Example \ref{ex2}, we decrease the signal strength, and compare different methods for the case where the position of the signals are randomly assigned. In Examples \ref{diff-nc} and \ref{diff-b}, performances of different $n_c$'s and $b$'s are reported, respectively. 

\begin{example}{(Different Correlation Settings)}\label{ex1}
We set $(n, p) = (300, 1000)$ and $\bbeta \in \mathbb{R}^{p}$ with the first 8 coordinates $(4, 3, 2, 0, 0, -4, 3, -2)$ and 0 elsewhere. For $i=1,\cdots, n$, we generate the response $y_i$ as follows
\[
y_i=\bx_i'\bbeta+\veps_i,
\]
where $\bx_i\stackrel{i.i.d.}{\sim} N(\bzero_p,\bSigma)$ with $\bzero_p$  the length-$p$ vector with 0 entries and $\veps_i\stackrel{i.i.d.}{\sim} N(0,1)$. The following three different correlation settings are considered.
\begin{enumerate}
\item[(a)] (Independent) $\Sigma_{j, k} = \mathbbm{1}\{j = k\}$.
\item[(b)] (Exponential Decay) $\Sigma_{j,k}=\rho^{|j-k|}$, with $\rho=0.2, 0.5$ and 0.7.
\item[(c)] (Equal Correlation) $\Sigma_{j,k}=\rho+(1-\rho)1\{j=k\}$, with $\rho=0.2, 0.5$ and 0.7.
\end{enumerate}
We repeat the simulation for 100 times.

\end{example}

\begin{figure}
\centering
\includegraphics[scale = 0.85]{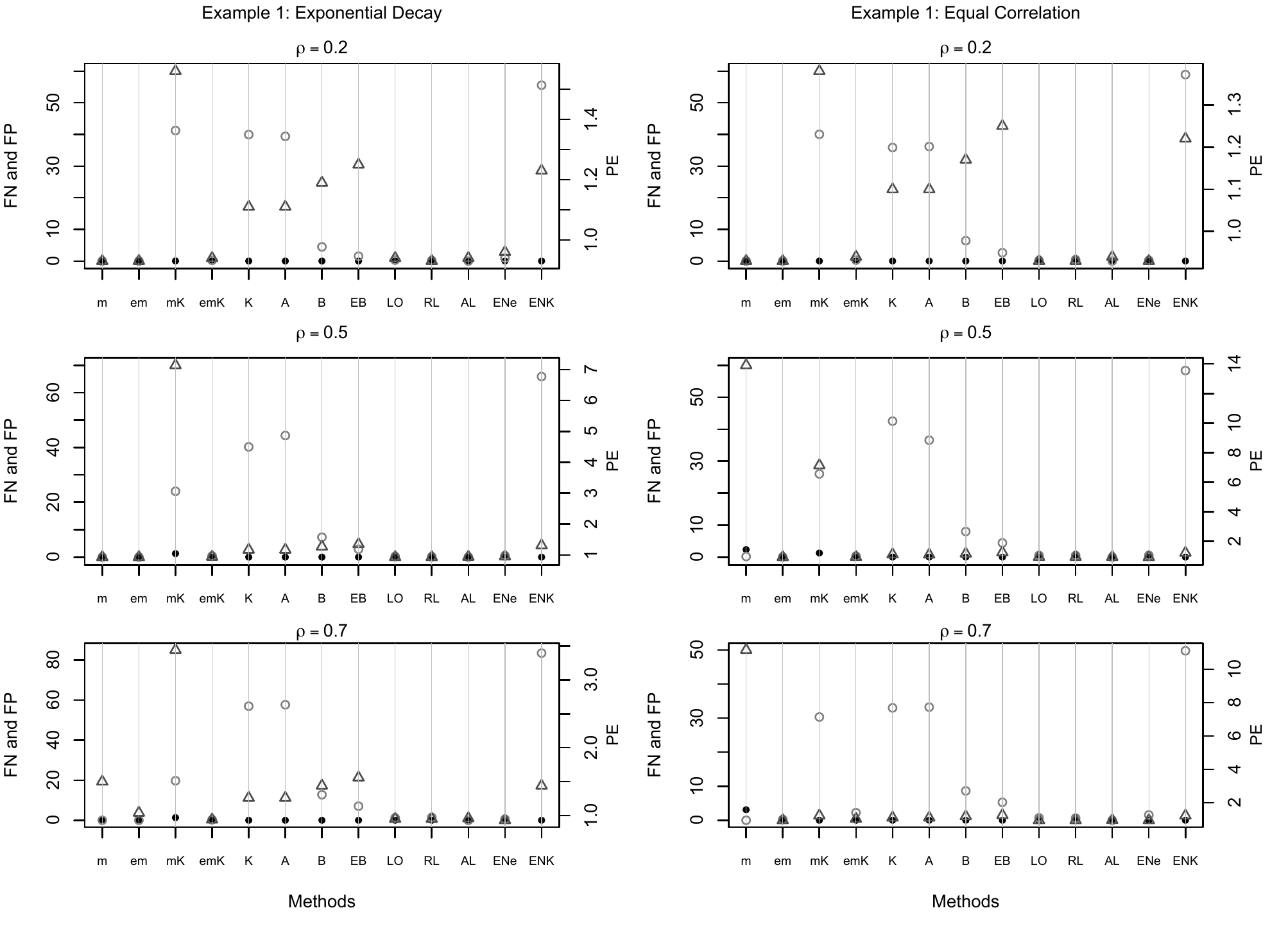}
\caption{Comparisons of different methods and scenarios in Example 1(b) and 1(c), which are presented in the left and right columns, respectively. Along the $x$-axis, from left to right, the methods are m-MCCV($n_v$), em-MCCV($n_v$), m-$K$-fold, em-$K$-fold, $K$-fold, AIC, BIC, EBIC, LARS-OLS, Relaxed Lasso, Elastic Net with em-MCCV($n_v$) and Elastic Net with $K$-fold. Means of FN, FP, and PE over 100 repetitions are labeled by symbols $\bullet$, \color{grey1} {\small $\triangle$}\color{black}, and \color{grey2} $\circ$\color{black}, respectively.}\label{fig-ex1b}
\end{figure}

The results for Example \ref{ex1} are reported in Table \ref{T1(a)} for case (a) and Figure \ref{fig-ex1b} for cases (b) and (c).  The detailed results for cases (b) and (c) are available in Table \ref{T1(b)} in the Appendix. We use the \pkg{glmnet} package to generate the Lasso solution paths for the whole dataset, and every subsample.  For modified Monte Carlo CV($n_v$) (m-MCCV($n_v$)) and exactly modified Monte Carlo CV($n_v$) (em-MCCV($n_v$)), we set $n_c = \lceil n^{3/4} \rceil = 73$, $n_v = n - n_c = 227$ 
and $b = 50$, which give robust results, with reasonable computation cost. The model selection performances of m-MCCV($n_v$), em-MCCV($n_v$), modified $K$-fold cross-validation (m-$K$-fold, $K = 10$), exactly modified $K$-fold cross-validation (em-$K$-fold, $K = 10$) and $K$-fold cross-validation ($K = 10$) in \pkg{glmnet}  package are presented, along with those of AIC, BIC, EBIC, LARS-OLS, relaxed Lasso\footnote{$\bhbeta({\lam})_{\text{relaxed Lasso}} \equiv \arg\min_{\bbeta} \frac{1}{2n} \|\by-\bX\bbeta\|^2 + \phi\lambda \|\bbeta\|_1$.}, adaptive Lasso \citep{Zou2006}, and Elastic Net. Both LARS-OLS and relaxed Lasso are computed by \pkg{R} package \pkg{relaxo}, adaptive Lasso solutions are obtained by \pkg{parcor}, with $10$-fold CV as the default tuning parameter selection method. In Figure \ref{fig-ex1b}, Tables \ref{T1(a)} and \ref{T1(b)}, results for em-MCCV($n_v$) applied to Elastic Net estimators\footnote{$\bhbeta({\lam})_{\text{e-net}} \equiv \arg\min_{\bbeta} \frac{1}{2n} \|\by-\bX\bbeta\|^2 + \alpha\lambda \|\bbeta\|^2 + (1-\alpha)\lambda \|\bbeta\|_1$,
where $\alpha = 0.5$ is used here.} are also included, whose paths are generated by \pkg{R} package \pkg{glmnet}, and $n_c = \lceil n^{2/3} \rceil$.  To compare the performance of different methods,  we report false negative (FN), {false positive} (FP), and  {prediction error} (PE). Here, PE is defined as the average squared prediction error calculated on an independent test dataset of size $n$. 


\begin{table}
\caption{Comparison of the performance for different methods, for the setting of Example \ref{ex1}(a). Results are reported in the form of mean (standard deviation). For (e)m-MCCV($n_v$), $n_c  = \lceil n^{3/4} \rceil$ and $b = 50$. For m-$K$-fold and $K$-fold, $K = 10$.}\label{T1(a)}
\centering
\begin{tabular}[h]{l | rrr}
 \hline
Methods &  FN & FP & PE \\
\hline
& \multicolumn{3}{c}{Independent} \\
\hline
m-MCCV($n_v$) &  0.00(0.00) & 0.01\,~(0.10) 
& 0.93(0.02) \\
em-MCCV($n_v$) & 0.00(0.00) & 0.00\,~(0.00) 
&0.93(0.01) \\
m-$K$-fold &  0.00(0.00) & 38.22(16.19) 
&1.34(0.18) \\
em-$K$-fold & 0.00(0.00) & 0.38\,~(1.45) 
&0.94(0.05) \\
$K$-fold & 0.00(0.00) & 34.99(22.06) & 1.11(0.06) \\
AIC & 0.00(0.00) & 35.32(18.90) & 1.11(0.06)\\
BIC & 0.00(0.00) & 4.47\,~(3.10) & 1.17(0.08) \\
EBIC & 0.00(0.00) & 1.37\,~(1.40) & 1.22(0.09) \\
LARS-OLS & 0.00(0.00) & 0.26\,~(0.80) & 0.94(0.04)\\
Relaxed Lasso & 0.00(0.00) & 0.34\,~(1.44) & 0.94(0.04) \\
AdaLasso & 0.00(0.00) & 0.00\,~(0.00) & 0.94(0.02)\\
\hline
Elastic Net & \\
\hline
 em-MCCV($n_v$) &  0.00(0.00) & 0.87\,~(1.45) & 0.95(0.05)\\
$K$-fold &  0.00(0.00) & 54.78(25.23) & 1.21(0.08)\\
\hline
\end{tabular}
\end{table}

For the independent design case, the two CV($n_v$) methods have no false negative and almost no false positive, which indicates that they nearly achieve the model selection consistency. On the other hand, all the other cross-validation methods have a significantly large number of false positives. It is interesting to note that em-$K$-fold has similar behavior as LARS-OLS, as expected. But still, they have larger FP than that of em-MCCV($n_v$). And the PEs of the m-MCCV and em-MCCV are  smaller than those of the other methods.


The results for AIC, BIC and EBIC are also reported. Notice that EBIC has the best performance among the three information criterion based methods, although it is still worse than the em-MCCV. One advantage of the information criterion based methods is that they are more efficient to compute without the need of cross-validation. As suggested by one referee, we also include the comparison with adaptive Lasso, which turns out to have a comparable  performance with the EMCC-based Lasso. We would like to point out that the main goal here is to find a better tuning parameter selection method for the Lasso estimator, while adaptive Lasso has different solutions from Lasso. We refer the interested readers to \cite{Zou2006} for a detailed treatment of adaptive Lasso with the advantages over the ordinary Lasso. 

For the exponential decay cases, m-MCCV and em-MCCV still perform very well with the smallest FP and FN among all the different methods, especially compared with more than 40 false positives on average in $K$-fold CV ($K = 10$). Due to the correlation in design matrices, compared to uncorrelated cases, the m-$K$-fold method miss some important variables while the em-$K$-fold can pick them up.

To make the case more extreme, and to see the difference between two modified criteria, we report the results for the equal correlation cases, which are not common in real applications. From the results, we can see as an approximation, the m-MCCV method is too aggressive and select too few variables (it misses some important variables) when there is strong correlation among variables. In the presence of strong correlation, we recommend using the em-MCCV, which has superior performance.

Now, we study the computation cost of different methods. It is well known that the cost for calculating the solution path of Lasso is $O(np\min\{n,p\})$ \citep{EfronHJT2007, Meinshausen}. The information type methods, including AIC, BIC and EBIC, have the best performance, since they only need one-time calculation for each model on the solution path, which leads to the computation cost $O(np\min\{n,p\})$. In Table \ref{table-cost}, we show the computing cost comparison for other methods for the purpose of choosing the tuning parameter.   We see that the (e)m-$K$-fold and LARS-OLS have computation cost of the same order. Since relaxed Lasso involves cross-validation on a 2-dimensional parameter grid, the computation cost is $O(KLnp\min\{n,p\})$, where $L$ is the number of different $\phi$'s, representing different level of penalty on the specific variable. Depending on the values of $b$ and $L$, we expect (e)m-MCCV and relaxed Lasso have similar computation cost. 

\begin{table}[t]\caption{Computation cost comparison for the tuning parameter selection. Here, $K$ is the number of folds, $b$ is the number of splits in (e)m-MCCV and $L$ is the number of different $\phi$'s considered in the relaxed Lasso. 
\label{table-cost}
}
\centering
\begin{tabular}[h]{l|l|l|l}
\hline
(e)m-$K$-fold & LARS-OLS & (e)m-MCCV & relaxed Lasso\\
\hline
 $O(Knp\min\{n,p\})$&$O(Knp\min\{n,p\})$&$O(bnp\min\{n,p\})$&$O(KLnp\min\{n,p\})$\\
 \hline
 \end{tabular}
\end{table}

As mentioned in Section \ref{sec-mcc}, the idea of removing the systematic bias can be easily extended to other popular penalties. As expected, em-MCCV($n_v$) leads to much smaller FPs than those of $K$-fold CV for Elastic Net, and it also leads to a smaller PE.

\begin{example}{(Random Position with Small Signals)}\label{ex2}
We use exactly the same setting as Example \ref{ex1} except to reduce the signal strength by setting the $\bbeta$ to have the non-zero coordinates $(1.2, 0.8,$ $0.4)$. In this example, we use independent design and exponential decay design with $\rho = 0.5$. In the exponential decay design case, aside from the case where the signals lie in the first three coordinates, we also consider the case when the signals are randomly positioned. \end{example}

The results of Example \ref{ex2} are reported in Figure \ref{fig-ex2}, which is left in the Appendix. The purpose of this example is to investigate the performances of different methods when the signals are of small strength, also to show the results when the signals are randomly positioned. In general, we have similar conclusions as Example \ref{ex1}. It is interesting to notice that when the signals are randomly positioned, the methods have similar behavior as independent case, which is due to the fact that exponential decay correlation implies the signal variables are approximately independent since their positions can be very different.

\begin{example}\label{diff-nc} {(Different $n_c$)}
In this example, we would like to study the influence of different splitting rates for the construction and validation dataset (i.e., different values of $n_c$). We report the results for independent and exponential decay settings in Example 1 with $\rho = 0.5$.  \end{example}

\begin{figure}
\centering
\subfigure[]{
\includegraphics[scale=0.4]{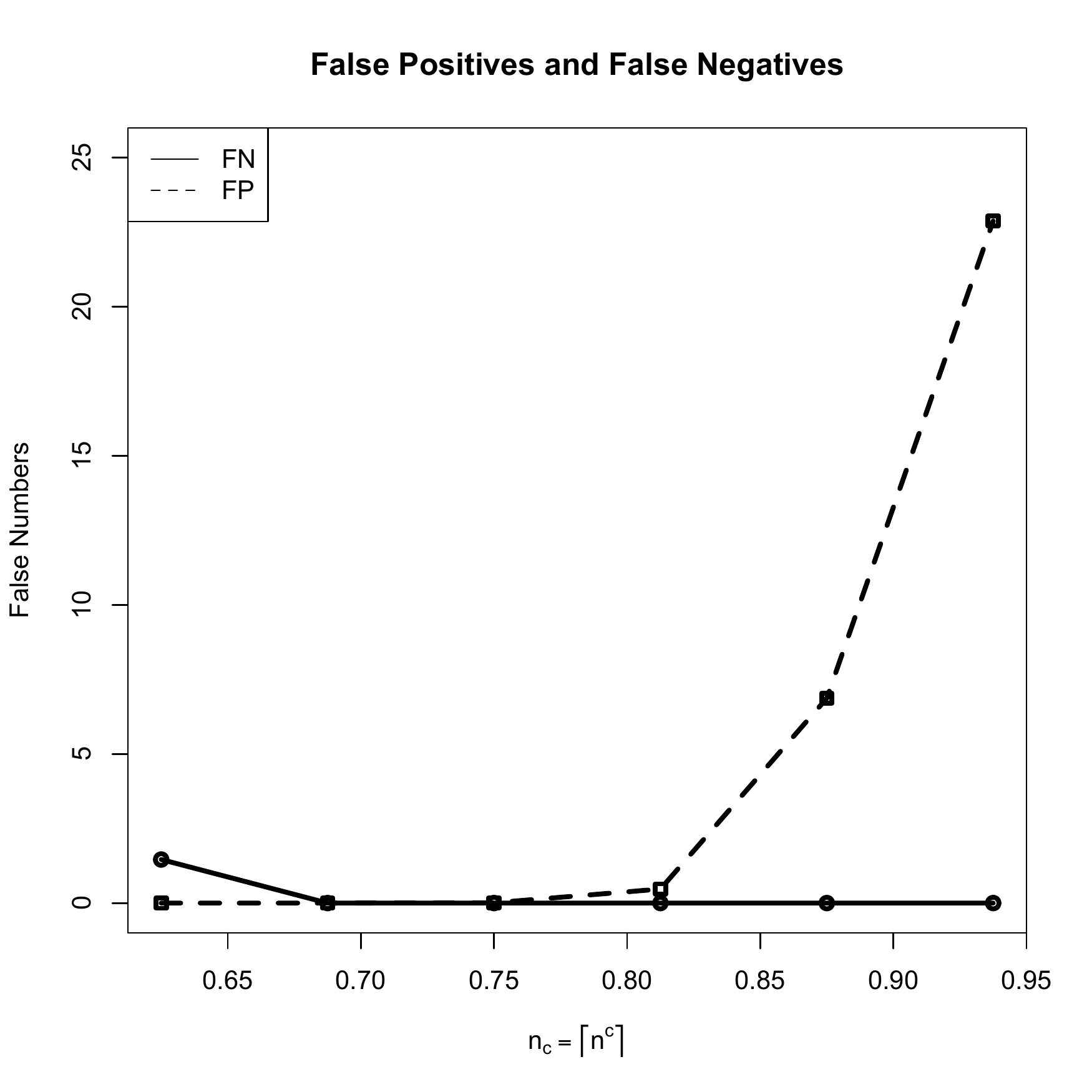}
\label{m-0}
}
\subfigure[]{
\includegraphics[scale=0.4]{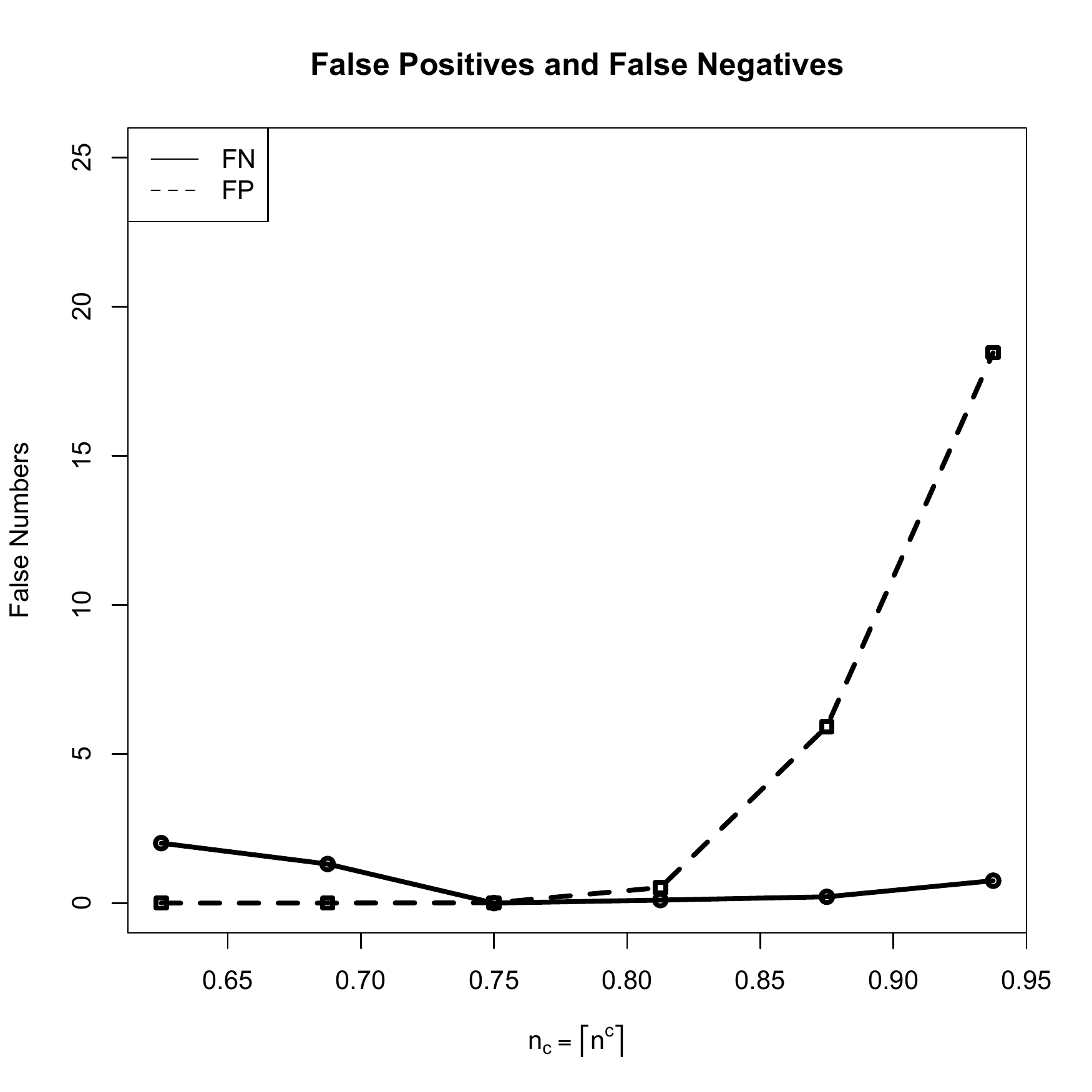}
\label{m-5}
}
\caption{FN and FP of m-MCCV($n_v$) for independent and exponential decay designs.}\label{fig-2}
\end{figure}

Figure \ref{fig-2} summarizes the results of m-MCCV($n_v$) and em-MCCV($n_v$) with $n_c$ varying from $\lceil n^{10/16} \rceil$ to $\lceil n^{15/16} \rceil$.
For both methods, $n_c = \lceil n^{12/16} \rceil $ leads to the best performance, in terms of FN, FP and PE. But if $n_c$ is smaller than $\lceil n^{12/16} \rceil$, (e.g., when $n_c=\lceil n^{10/16} \rceil$)  m-MCCV misses some important variables. The possible reason is that when $n_c$ is too small in this fixed sample simulation example, the penalty imposed on selecting one more variable exceeds the loss of missing one important variable.
For $c = 14/16$ and $c = 15/16$ cases, the approximation results in substantial number of FN, which the em-MCCV($n_v$) can remedy. If  computation cost is an issue,  m-MCCV($n_v$) is preferred as long as $n_c$ is properly chosen and the correlation is not too extreme.

It is also interesting to see the trend in false positive. When $n_c$ increases, the false positive also increases. This is consistent with our intuition, that if the validation dataset is small, detection of different models becomes more difficult. It is clear to see, along with the increasing of $n_c$ rate, i.e., the ratio of constructing sample size to the whole sample size, there will be more false positives  and  less false negatives.  This also gives us a rough guideline how to choose the proper $n_c$ rate. From the figures we can see, if $c = 3/4$, both false positives and false negatives attain a relatively low value. Larger or smaller $c$ value will cause one side uneven.

\begin{example}\label{diff-b} {(Different Number of Splits $b$)}
For the setting used in Example \ref{ex1}, we vary $b$ from 10 to 300, which is equivalent to the order $n^{3/8}$ to $n$. In Table \ref{T-b}, we show that results for Modified-Reversed-$K$-fold (m-r-$K$-fold) strategy, i.e., instead of using one fold to validate and $K-1$ folds to construct, we use $K-1$ folds to validate and one fold to construct. \end{example}

\begin{table}
\caption{Reversed $K$-fold splitting strategy, with $K = 10$. Results are reported in the form of mean(standard deviation). (e)m-r-$K$-fold CV is short for (exactly) modified reversed $K$-fold cross-validation. }\label{T-b}
\centering
\begin{tabular}[h]{l | rrrr }
\hline		
Methods &  FN & FP & PE \\
\hline
 \multicolumn{4}{c}{Independent}\\
\hline
m-r-$K$-fold CV &  0.00(0.00) & 1.04\,~(0.06) & 1.33(0.38)  \\
em-r-$K$-fold CV&  0.00(0.00) & 0.10\,~(0.10) & 1.06(0.02)\\
$K$-fold CV & 0.00(0.00) & 34.99(22.06) & 1.11(0.06)\\
\hline
\multicolumn{4}{c}{Exponential Decay ($\rho = 0.5$)}\\
\hline
m-r-$K$-fold CV&  0.00(0.00) & 0.98\,~(0.04) & 0.17(0.38)  \\
em-r-$K$-fold CV &  0.00(0.00) & 0.06\,~(0.03) & 0.95(0.05)\\
$K$-fold CV & 0.00(0.00) & 40.21(18.18) & 1.17(0.07) \\
\hline
\multicolumn{4}{c}{Equal Correlation ($\rho = 0.5$)}\\
\hline
m-r-$K$-fold CV&  0.00(0.00) & 1.02\,~(0.05) & 0.98(0.02)  \\
em-r-$K$-fold CV &  0.21(0.40) & 4.27\,~(5.37) & 0.99(0.05)\\
$K$-fold CV & 0.00(0.00) & 42.52(22.59) & 1.11(0.06)  \\
\hline
\end{tabular}
\end{table}

The simulation results for different $b$ are almost the same in our experiments, which indicates that FN, FP and PE are not sensitive to the choice of $b$ in both correlation settings. The detailed results can be found in Table \ref{T3}. In real applications, a conservative way is to set $b$ slightly larger if enough computational resource is available. In all the other simulations and the following real data analysis, we use $b = 50$, which exceeds $n^{1/2}$ and produces stable results.

In Table \ref{T-b}, it is surprising to see, by using a small number of splits (for example $K=10$), the Modified-Reversed-$K$-fold strategy can achieve very good results. This strategy guarantees each sample appears in the construction/validation set for the same number of times. It is worth to point out that although the results are very good when using the Modified-Reversed-$K$-fold, this splitting strategy does not belong to the block incomplete design or BICV \citep{Shao1993}, since it does not balance the frequency of the pairs. In addition, the Modified-Reversed-$K$-fold takes less time to compute and has the same order of computation cost as the regular $K$-fold CV.

\section{Data Analysis}\label{sec-data}

We  now illustrate one application of the proposed m-MCCV($n_v$) method via the dataset reported by \cite{Scheetz2006} and analyzed by \cite{HuangEtal2010} and \cite{FanEtal2011}. In this dataset, for harvesting of tissue from the eyes and subsequent microarray analysis, 120 12-week-old male rats were selected. The microarrays used to analyze the RNA from the eyes of these animals contain more than 31,042 different probe sets (Affymetric GeneChip Rat Genome 230 2.0 Array). The intensity values were normalized using the robust multichip averaging method \citep{Irizarry2003} to obtain summary expression values for each probe set. Gene expression levels were analyzed on a logarithmic scale.

Following \cite{HuangEtal2010} and \cite{FanEtal2011}, we are interested in finding the genes that are related to the TRIM32 gene, which was recently found to cause Bardet-Biedl syndrome \citep{Chiang2006} and is a genetically heterogeneous disease of multiple organ systems, including the retina. Although more than 30,000 probe sets are represented on the Rat Genome 230 2.0 Array, many of these are not expressed in the eye tissue. We only focus  on the 18,975 probes that are expressed in the eye tissue.

We use \pkg{R} package \pkg{glmnet} to compute the Lasso solution paths, and compare our proposed modified CV criterion with the 10-fold CV. The results are presented in Table \ref{Data}, with $n_c = \lceil n^{3/4} \rceil$ and $b = 50$. We can achieve the same PE with only 17 variables on average, compared to 60 in 10-fold CV case. This shows the m-MCCV($n_v$) can generate more a parsimonious model while keeping the same prediction power, which could be potentially helpful in guiding the biologists to focus on the fewer selected genes. 

\begin{figure}
\centering
\includegraphics[scale = 0.8]{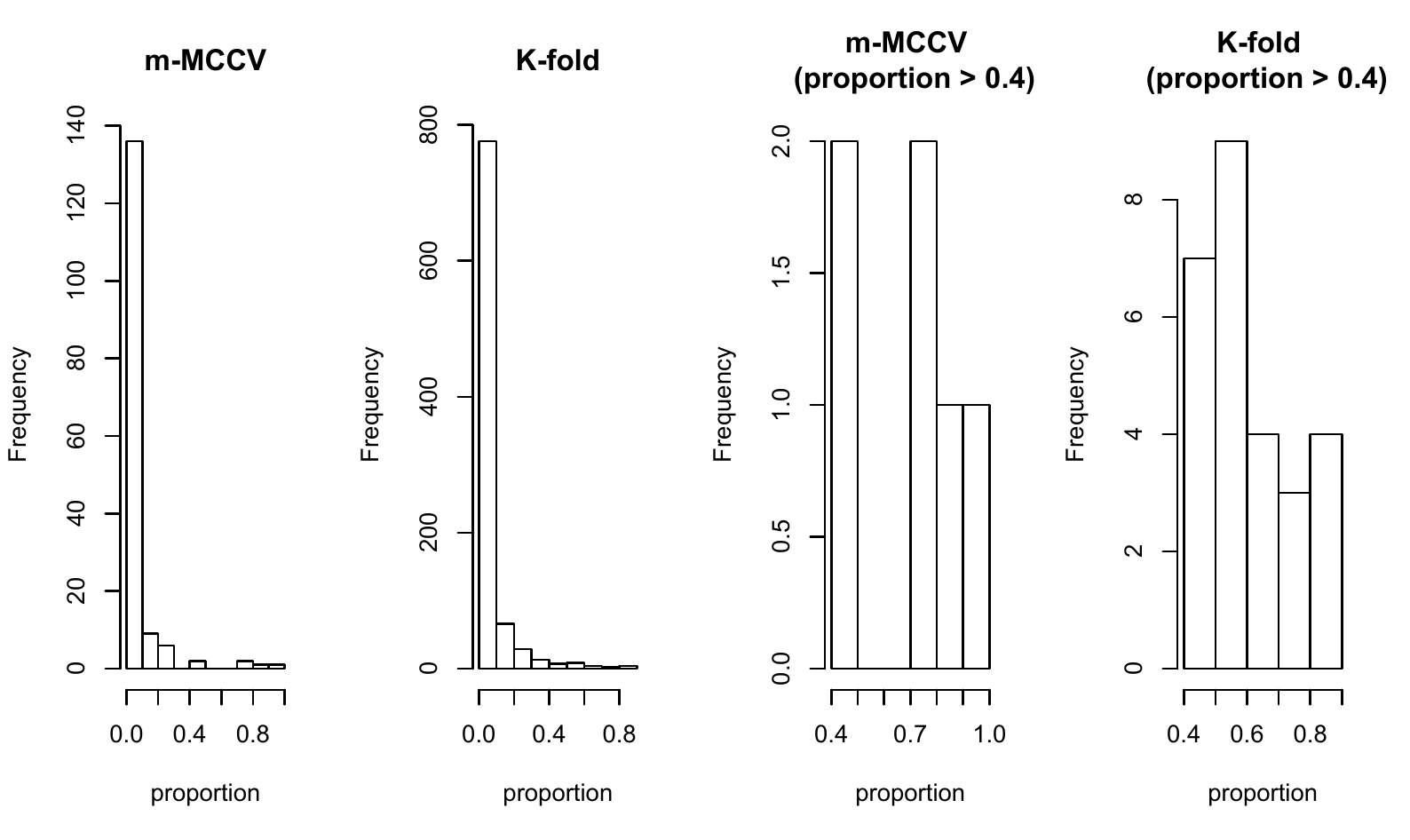}

\caption{Histograms of gene appearance proportions using m-MCCV($n_v$) and $K$-fold CV. The two left figures are the histograms of all the genes appeared in 100 repetitions and two right ones are genes with proportion greater than $0.4$ only.\label{real-data-freq}}
\end{figure}

In Figure \ref{real-data-freq}, we show the histograms of the proportion of the gene being selected in 100 splits.  In the stability selection theory developed in \cite{MeinshausenBuehlmann2010}, the selection proportion of a certain variable can represent the degree of ``stability" for the Lasso estimator. As a result, the variables with larger values of proportion are more likely to be ``important". We reproduce the histogram in the right two subfigures of Figure \ref{real-data-freq} for genes with selected proportion larger than 0.4. It is worth noting that the histogram of proposed m-MCCV has a big gap between 0.5 and 0.7,  while no similar pattern is observed for that of $K$-fold CV. This particular gap may serve as a natural threshold as whether the gene is important. 

In Table \ref{table-frequencies}, we list the symbols of the genes selected by each method along with the selection proportion up to the median number of selected variables in 100 splits.  Note that ``---" represents there is no known symbol for the corresponding gene. In general, genes with symbols have been shown to carry certain biological functions. It is observed that 70.6\% of the genes selected by m-MCCV($n_v$) have gene symbols, compared with 59.3\% for 10-fold CV. In addition, if we adopt 0.6 as a selection cut-point (this corresponds to the gap mentioned in (b)), all the four genes m-MCCV($n_v$) selected have gene symbols, compared with only 50\% for the 10-fold CV.

\begin{table}
\caption{Comparison results of m-MCCV($n_v$) and 10-fold-CV for real dataset, model size, PE and RPE are presented in the form of mean(standard deviation), 100 repetitions are conducted.}\label{Data}
\centering
\begin{tabular}[h]{l l | rr }
\hline
\multicolumn{2}{c|}{m-MCCV($n_v$)}  & \multicolumn{2}{c}{10-fold CV} \\
 \hline
size & PE & size & PE \\
\hline
17.90(2.99) & 0.01(0.01) & 60.30(17.07) & 0.01(0.00) \\
\hline
\end{tabular}
\end{table}

\section{Discussion}\label{sec-discussion}

In this paper, we systematically investigated the behavior of different types of cross-validation, with different criterion functions, applied to  the tuning parameter selection problem in Lasso penalized linear regression models. By removing the bias caused by the Lasso penalty, we proposed a new  cross-validation method em-MCCV with an approximated version m-MCCV. Both methods work well in simulations and real applications.

Some interesting future work includes the theoretical investigation of the inconsistency of the traditional $K$-fold cross-validation. Also, we conjecture that the newly proposed em-MCCV($n_v$) is model selection consistent under certain technical conditions. Other work includes extensions to generalized linear models, Cox models and semi-parametric models. 

\section*{Appendix}
We include the detailed tables and figures of Sections 4 and 5 in the appendix. 
\begin{sidewaystable}
\caption{Comparison of the performance for different methods, for exponential decay cases with different $\rho$ in Example \ref{ex1}(b) and (c). Results are reported in the form of mean (standard deviation). For (e)m-MCCV($n_v$), $n_c  = \lceil n^{3/4} \rceil$ and $b = 50$. For m-$K$-fold and $K$-fold, $K = 10$.}\label{T1(b)}
\centering
{\footnotesize
\begin{tabular}[h]{l | rrr | rrr | rrr}
 \hline
Methods &  FN & FP & PE &  FN & FP & PE &  FN & FP & PE \\
\hline
Exponential Decay & \multicolumn{3}{c|}{$\rho = 0.2$} & \multicolumn{3}{c|}{$\rho = 0.5$} & \multicolumn{3}{c}{$\rho = 0.7$}\\
\hline
m-MCCV($n_v$) & 0.00(0.00) & 0.00\,~(0.00) &0.93(0.01) & 0.00(0.00) & 0.01\,~(0.10) &0.93(0.02) & 0.26(0.09) & 0.01\,~(0.10) &1.50(0.17) \\
em-MCCV($n_v$) & 0.00(0.00) & 0.00\,~(0.00) &0.93(0.01) & 0.00(0.00) & 0.03\,~(0.17) &0.93(0.02) & 0.08(0.09) & 0.00\,~(0.00) &1.04(0.32) \\
m-$K$-fold &  0.03(0.30) & 41.24(10.34) &1.56(1.57) & 1.28(1.73) & 24.02(24.10) &7.14(8.25) &  1.32(0.10) & 19.73(10.20) &3.44(5.21)\\
em-$K$-fold & 0.00(0.00) & 0.23\,~(0.93) &0.94(0.04)& 0.00(0.00) & 0.42\,~(1.43)
&0.94(0.04) & 0.00(0.00) & 0.25\,~(0.70) &0.94(0.03)\\
$K$-fold & 0.00(0.00) & 39.93(29.03) & 1.11(0.06) & 0.00(0.00) & 40.21(18.18) & 1.17(0.07)  & 0.00(0.00) & 56.93\,~(5.73) & 1.26(0.09) \\
AIC &  0.00(0.00) & 39.39(18.01) & 1.11(0.06) & 0.00(0.00) & 44.36(20.54) & 1.17(0.07) &  0.00(0.00) & 57.65(17.81) & 1.26(0.09) \\
BIC & 0.00(0.00) & 4.48\,~(3.42) & 1.19(0.08)  & 0.00(0.00) & 7.25\,~(4.49) & 1.27(0.10) & 0.00(0.00) & 12.76\,~(5.65) & 1.44(0.12) \\
EBIC & 0.00(0.00) & 1.52\,~(1.53) & 1.25(0.10) & 0.00(0.00) & 2.90\,~(2.23) & 1.35(0.11)  & 0.00(0.00) & 7.02\,~(3.33) & 1.56(0.16) \\
LARS-OLS & 0.00(0.00) & 0.29\,~(1.12) & 0.94(0.04) & 0.00(0.00) & 0.23\,~(0.71) & 0.93(0.03) & 0.00(0.00) & 1.19\,~(1.13) & 0.95(0.03)\\
Relaxed Lasso & 0.00(0.00) & 0.16\,~(0.72) & 0.93(0.03)  & 0.00(0.00) & 0.15\,~(0.61) & 0.93(0.03) & 0.00(0.00) & 1.41\,~(1.27) & 0.95(0.04)\\
AdaLasso & 0.00(0.00) & 0.00\,~(0.00) & 0.94(0.02) & 0.00(0.00) & 0.00\,~(0.00) & 0.94(0.02)  & 0.00(0.00) & 0.00\,~(0.00) & 0.96(0.03)\\
\hline
Elastic Net & \\
\hline
em-MCCV($n_v$) &  0.00(0.00) & 0.95\,~(1.53) & 0.96(0.06) &  0.00(0.00) & 0.53\,~(1.03) & 0.95(0.04)  &  0.00(0.00) & 0.54\,~(0.82) & 0.93(0.02)\\
$K$-fold &  0.00(0.00) & 55.56(22.67) & 1.23(0.08)  &  0.00(0.00) & 65.87(21.73) & 1.31(0.09)  &  0.00(0.00) & 83.45(16.91) & 1.44(0.13)\\
\hline\hline
Equal Correlation & \multicolumn{3}{c}{$\rho = 0.2$} & \multicolumn{3}{c}{$\rho = 0.5$} & \multicolumn{3}{c}{$\rho = 0.7$}\\
\hline
m-MCCV($n_v$) &  0.00(0.00) & 0.00\,~(0.00) &0.93(0.02)& 2.36(1.22) & 0.20\,~(0.78) 
&13.91(6.54)  &  3.09(0.47) & 0.00\,~(0.00) 
&11.16(0.67) \\
em-MCCV($n_v$) &  0.00(0.00) & 0.00\,~(0.00) &0.93(0.02) & 0.00(0.00) & 0.06\,~(0.34) 
&0.93(0.02) &  0.00(0.00) & 0.22\,~(0.52) 
&0.93(0.02)\\
m-$K$-fold & 0.00(0.00) & 40.04\,~(8.19) &1.38(0.11)& 1.28(1.74) & 26.03(21.32) 
&7.14(8.25)&  0.00(0.00) & 30.31(10.53) 
&1.22(0.20)\\
em-$K$-fold & 0.00(0.00) & 0.29\,~(1.00) &0.94(0.04)& 0.00(0.00) & 0.30\,~(0.79) 
&0.94(0.03) &  0.00(0.00) & 2.23\,~(0.07) 
&1.03(0.02)\\
$K$-fold &  0.00(0.00) & 35.86(18.53) & 1.10(0.06)& 0.00(0.00) & 42.52(22.59) & 1.11(0.06) & 0.00(0.00) & 32.99(14.74) & 1.10(0.06)  \\
AIC & 0.00(0.00) & 36.18(17.37) & 1.10(0.06)  & 0.00(0.00) & 36.58(16.38) & 1.10(0.06)  & 0.00(0.00) & 33.21(12.06) & 1.10(0.06) \\
BIC & 0.00(0.00) & 6.43\,~(3.82) & 1.17(0.08) & 0.00(0.00) & 8.01\,~(3.95) & 1.17(0.07) & 0.00(0.00) & 8.63\,~(4.27) & 1.18(0.07)\\
EBIC & 0.00(0.00) & 2.63\,~(2.05) & 1.25(0.11) & 0.00(0.00) & 4.44\,~(3.47) & 1.25(0.11) & 0.00(0.00) & 5.28\,~(3.61) & 1.25(0.10)\\
LARS-OLS & 0.00(0.00) & 0.21\,~(0.76) & 0.93(0.03)  & 0.00(0.00) & 0.51\,~(1.51) & 0.94(0.05) & 0.00(0.00) & 0.73\,~(1.25) & 0.94(0.04)\\
Relaxed Lasso & 0.00(0.00) & 0.43\,~(1.96) & 0.93(0.04) & 0.00(0.00) & 0.51\,~(1.37) & 0.94(0.03)  & 0.00(0.00) & 0.63\,~(1.14) & 0.94(0.03)\\
AdaLasso & 0.00(0.00) & 0.02\,~(0.20) & 0.94(0.02) & 0.00(0.00) & 0.01\,~(0.10) & 0.94(0.02) & 0.00(0.00) & 0.01\,~(0.10) & 0.95(0.03)\\
\hline
Elastic Net & \\
\hline
 em-MCCV($n_v$) & 0.00(0.00) & 0.20\,~(0.60) & 0.93(0.03)  &  0.00(0.00) & 0.54\,~(0.91) & 0.93(0.02) &  0.00(0.00) & 1.56\,~(1.93) & 0.94(0.03)\\
$K$-fold &  0.00(0.00) & 58.90(23.36) & 1.22(0.08)  &  0.00(0.00) & 58.35(21.19) & 1.22(0.08) &  0.00(0.00) & 49.76(11.63) & 1.22(0.08)\\
\hline
\end{tabular}}
\end{sidewaystable}

\begin{figure}
\centering
\includegraphics[scale = 0.8]{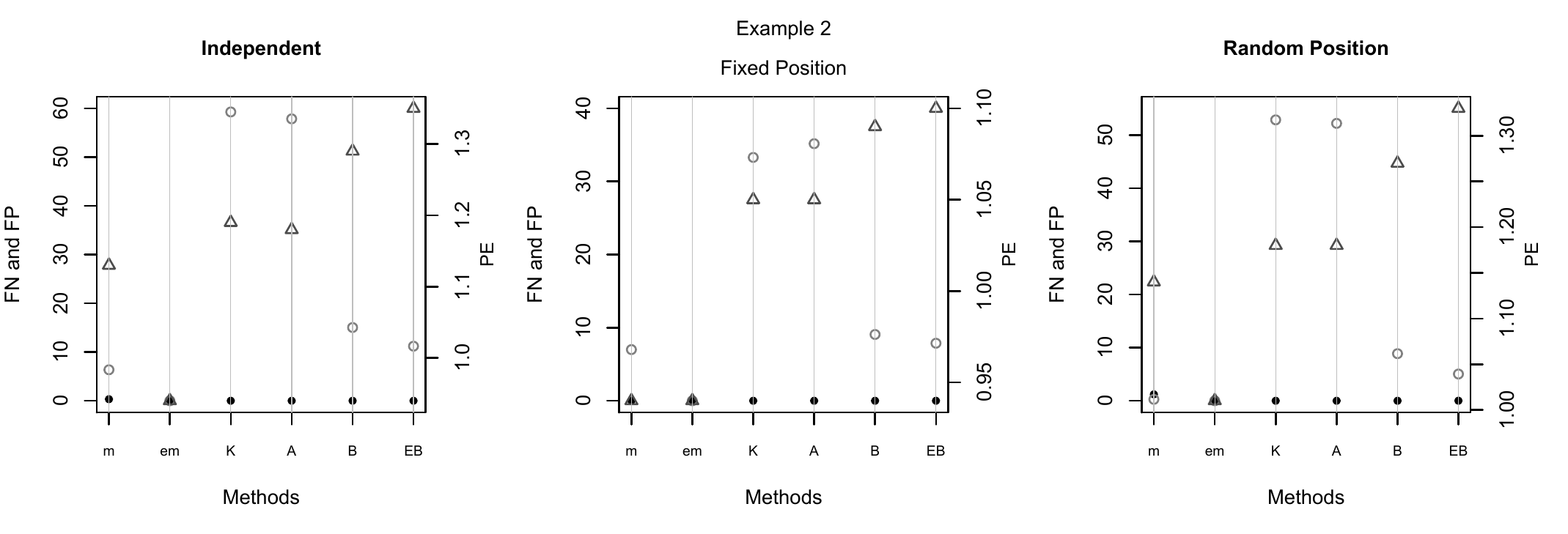}
\caption{Comparisons of different methods and scenarios in Example 2. Along the $x$-axis, the methods are m-MCCV($n_v$), em-MCCV($n_v$), $K$-fold, AIC, BIC and EBIC, sequentially. Means of FN, FP and PE over 100 repetitions are labeled by symbols $\bullet$, \color{grey1} {\small $\triangle$}\color{black}, and \color{grey2} $\circ$ \color{black}, respectively.}\label{fig-ex2}
\end{figure}

\begin{table}
\caption{Comparison of m-MCCV($n_v$) with different $b$, using the settings in Example \ref{ex1}, with $n_c = \lceil n^{3/4}\rceil$. Results are reported in the form of mean(standard deviation). }\label{T3}
\centering
\begin{tabular}[h]{l l | rrr }
 \hline
Methods & $b$ &  FN & FP & PE\\
\hline		
\multicolumn{5}{c}{Independent}\\
\hline
\multirow{4}{*}{m-MCCV($n_v$)} & $10$ &  0.00(0.00) & 0.01(0.10) & 0.92(0.02)  \\
& $50$ &  0.00(0.00) & 0.01(0.10) & 0.92(0.02)\\
& $150$ &  0.00(0.00) & 0.01(0.10) & 0.92(0.02)  \\
& $300$ &  0.00(0.00) & 0.01(0.10) & 0.92(0.02)\\
\hline
\multicolumn{2}{c|}{$K$-fold CV} & 0.00(0.00) & 34.99(22.06) & 1.33(0.16) \\
\hline
\multicolumn{5}{c}{Exponential Decay ($\rho = 0.5$)} \\
\hline
\multirow{4}{*}{m-MCCV($n_v$)} & $10$  & 0.00(0.00) & 0.01(0.10) &  0.93(0.02) \\
& $50$  & 0.00(0.00) & 0.01(0.10) &  0.93(0.02) \\
& $150$ & 0.00(0.00) & 0.01(0.10) & 0.93(0.02) \\
& $300$ & 0.00(0.00) & 0.01(0.10) & 0.93(0.02) \\
\hline
\multicolumn{2}{c|}{$K$-fold CV}  & 0.00(0.00) & 40.21(18.18) & 1.17(0.07)  \\
\hline
\end{tabular}
\end{table}

\begin{table}[!ht]
\caption{Proportions of gene being selected in 100 splits for m-MCCV($n_v$) and $K$-fold CV.}\label{table-frequencies}
\centering
\begin{tabular}[h]{l l || rr | rr | rr }
\hline
\multicolumn{2}{c||}{m-MCCV($n_v$)}  & \multicolumn{6}{c}{10-fold CV} \\
 \hline
Gene Symbol & prop. & Gene Symbol & prop. & Gene Symbol & prop. & Gene Symbol & prop. \\
\hline
TRIM41 & 0.95 & TRIM41 & 0.87 & CTDSPL & 0.50 & FRAS1 & 0.30\\
CCBL1 & 0.89 & TNFSF13 & 0.87 & RASL12 & 0.48 & RGD1307201 & 0.30\\
ES1 & 0.80 & --- & 0.87 & GJB2 & 0.47 & --- & 0.29\\
TRAK2 & 0.71 & --- & 0.87 & HERC3 & 0.46 & RGD1566403 & 0.29\\
--- & 0.42& --- & 0.75 & ASMT & 0.46 & WSB2 & 0.29\\
--- & 0.41  & --- & 0.74 & --- & 0.45 & RGD1308031 & 0.28\\
LOC678910 & 0.29  & --- & 0.71 & ADRB2 & 0.42 & --- & 0.28\\
RGD1305680 & 0.29  & ACAT1 & 0.70 & --- & 0.40 & --- & 0.28\\
--- & 0.27 & --- & 0.68 & --- & 0.39 & LOC296637 & 0.28\\
HDAC11 & 0.26 & ZFP367 & 0.68 & --- & 0.38 & CPNE9 & 0.28\\ 
--- & 0.25 &ANO10 & 0.62 & --- & 0.37 & --- & 0.27\\ 
TNFSF13 & 0.24 & FAM118B & 0.60 & --- & 0.36 & HINT1 & 0.27\\
--- & 0.19 & RGD1561792 & 0.58 & --- & 0.35 & RGD1309888 & 0.27\\
HEATR6 & 0.16 & PURB & 0.57 & --- & 0.35 & --- & 0.26\\
ACAT1 & 0.13 & --- & 0.56 & --- & 0.35 & BGLAP & 0.25\\
CABP1 & 0.13 & ACLY & 0.55 & --- & 0.34 & GFAP & 0.24\\  
MARVELD1 & 0.13 & HDAC11 & 0.55 & PRR12 & 0.34 & STK11 & 0.24\\
 & & --- & 0.55 & YTHDF3 & 0.33 & CYP4A3 & 0.24\\
 & & JAK2 & 0.53 & KLRD1 & 0.32 & ES1 & 0.24\\
  & & ATP6V1A & 0.52 & --- & 0.32 & \\ 
\hline
\hline
\end{tabular}
\end{table}

\section*{Acknowledgement}
The authors thank the editor, the associate editor, and two anonymous referees for their constructive comments which have greatly improve the scope of the paper.
\section*{Supplemental Materials}

\begin{description}


\item[\pkg{R} Code] The supplemental files for this article include \pkg{R} programs which can be used to replicate the simulation study and the real data analysis. The real data are available upon request. Please read file README contained in the zip file for more details. (MCC.zip) 

\end{description}

\bibliographystyle{spbasic}
\bibliography{CroVal}

\end{document}